\documentclass{raa}
\usepackage{graphicx,times}
\usepackage{natbib}
\usepackage{amssymb,amsmath}
\bibpunct{(}{)}{;}{a}{}{,}

\usepackage[a4paper=true,dvipdfm=true,pagebackref=true]{hyperref}
\hypersetup{pdftitle = The title of my PDF, pdfauthor = My name, pdfsubject= The subject, pdfkeywords = keyword1 keyword2 keyword3} 
\hypersetup{colorlinks = true, linkcolor = green, anchorcolor = red, citecolor = blue, filecolor = red, pagecolor = red, urlcolor = red}

\begin{document}

   \title{Statistical Analyses for NANOGrav 5-year Timing Residuals }

 \volnopage{ {\bf 2012} Vol.\ {\bf X} No. {\bf XX}, 000--000}
   \setcounter{page}{1}

   \author{Y. Wang\inst{1,2,3}, J.~M.~Cordes\inst{4}, F.~A.~Jenet\inst{2,3}, S.~Chatterjee
      \inst{4},  P.~B.~Demorest\inst{5}, T.~Dolch\inst{6}, J.~A.~Ellis\inst{7}, 
      M.~T.~Lam\inst{4}, D.~R.~Madison\inst{5}, M.~McLaughlin\inst{8},
      D.~Perrodin\inst{9}, J.~Rankin\inst{10}, X.~Siemens\inst{11}, 
      M.~Vallisneri\inst{7}
   }

   \institute{MOE Key Laboratory of Fundamental Physical Quantities Measurements, 
   School of Physics, Huazhong University of Science and Technology, 1037 Luoyu Road, 
   Wuhan, Hubei Province 430074, China; {\it ywang12@hust.edu.cn}\\
        \and
             Center for Advanced Radio Astronomy, University of Texas at Brownsville, 
             1 West University Boulevard, Brownsville, TX 78520, USA \\
	\and
Department of Physics and Astronomy, University of Texas at Brownsville, 
1 West University Boulevard, Brownsville, TX 78520, USA \\
\and 
Department of Astronomy, Cornell University, Ithaca, NY 14853, USA \\
\and
National Radio Astronomy Observatory, 520 Edgemont Road, Charlottesville, VA 22903, USA \\
\and
Department of Physics, Hillsdale College, 33 E. College Street, Hillsdale, MI 49242, USA \\
\and
Jet Propulsion Laboratory, California Institute of Technology, 4800 Oak Grove Drive, Pasadena, CA 91106, USA \\
\and
Department of Physics, West Virginia University, P.O. Box 6315, Morgantown, WV 26505, USA \\
\and
INAF-Osservatorio Astronomico di Cagliari, Via della Scienza 5, 09047 Selargius (CA), Italy \\
\and 
Department of Physics, University of Vermont, Burlington, VT 05405, USA \\
\and 
Center for Gravitation, Cosmology and Astrophysics, Department of Physics, University of Wisconsin-Milwaukee, P.O. Box 413, Milwaukee, WI 53201, USA \\
\vs \no
   {\small Received 2012 June 12; accepted 2012 July 27}
}

\abstract{
In pulsar timing, timing residuals are the differences between
the observed times of arrival and the predictions from the timing model. 
A comprehensive timing model will produce featureless
residuals, which are presumably composed of dominating noise and weak physical effects
excluded from the timing model (e.g. gravitational waves).
In order to apply the optimal statistical methods for detecting the weak 
gravitational wave signals,
we need to know the statistical properties of the noise components in the residuals.
In this paper we utilize a variety of non-parametric statistical tests to analyze 
the whiteness and Gaussianity 
of the North American Nanohertz Observatory for Gravitational Waves (NANOGrav) 5-year
timing data which are obtained from the Arecibo Observatory and the Green
Bank Telescope from 2005 to 2010 \citep{2013ApJ...762...94D}.
We find that most of the data are consistent with white noise; Many
data deviate from Gaussianity at different levels, nevertheless, 
removing outliers in some pulsars will mitigate the deviations.
}

   \authorrunning{Y. Wang et al. }            
   \titlerunning{Statistical Analyses for NANOGrav 5-year Timing Residuals}  
   \maketitle

%
\section{Introduction}           
\label{sect:intro}

Pulsar timing as a powerful technique has achieved many of the most
important science results in the pulsar astronomy. Timing of single
pulsars has been used as a probe of the dispersive interstellar medium
\citep{2002astro.ph..7156C}, to test the theories of gravitation
in the strong field regime \citep{1992PhRvD..45.1840D, 2003LRR.....6....5S,
2006Sci...314...97K}, to discover the first extraplanet system
\citep{1992Natur.355..145W}, to constrain the nuclear equation of state
of neutron star \citep{2010Natur.467.1081D, 2007PhR...442..109L, 2010arXiv1012.3208L}.
It has provided the first evidence of the existence of gravitational waves
\citep{1982ApJ...253..908T, 1989ApJ...345..434T}. Timing a number of pulsars and
analyzing the data coherently can be used to search for the irregularity of terrestrial
time standard and develop a time-scale based on 
pulsars \citep{2012MNRAS.427.2780H}, and to deepen our understanding 
of solar system dynamics \citep{2010ApJ...720L.201C}. Amazingly it can be
operated as a galactic scale detector for very-low frequency gravitational
waves \citep{1983ApJ...265L..39H, 1990ApJ...361..300F, 2005ApJ...625L.123J}.

Pulsar timing array (PTA) is an experiment to regularly observe a set of 
millisecond pulsars (MSPs). Currently, three PTAs, i.e., 
the North American Nanohertz Observatory for Gravitational
Waves (NANOGrav, \citet{2013ApJ...762...94D}), 
the Parkes Pulsar Timing Array (PPTA, \citet{2013PASA...30...17M}) 
and the European 
Pulsar Timing Array (EPTA, \citet{2010CQGra..27h4014F}), have 
started to produce astrophysically important results. 
These PTAs compose the
International Pulsar Timing Array (IPTA, \citet{2013arXiv1309.7392M,2014arXiv1409.4579M}) with
approximately 50 pulsars regularly monitored. The first data combination has 
been released \citep{2016MNRAS.458.1267V}.

The PTA is sensitive to the very low frequency ($10^{-9}$--$10^{-7}$ Hz)
gravitational waves (GWs), which is complementary to the ground based
interferometric detectors (e.g.,  LIGO \citep{2009RPPh...72g6901A} and
Virgo \citep{2011CQGra..28k4002A}) running in the high frequency
band ($10 -10^3$ Hz), and the space based laser rangers (e.g., eLISA 
\citep{2013arXiv1305.5720C} and TianQin \citep{2016CQGra..33c5010L}) 
proposed for the low frequency band (0.1 mHz--0.1 Hz).
Potential sources of GWs in the very low frequencies include the supermassive
black hole binaries \citep{2003ApJ...583..616J, 2003ApJ...590..691W,
2008MNRAS.390..192S}, 
the cosmic strings \citep{2005PhRvD..71f3510D, 2010PhRvD..81j4028O},
and the relic gravitational waves \citep{2005PhyU...48.1235G}.

At the current timing precision, it is very likely that the noises of different kinds 
are the dominating components of the timing residuals \citep{2006ApJ...653.1571J,
2011MNRAS.414.3117V, 2013ApJ...762...94D, 2013Sci...342..334S, 2014ApJ...794..141A}.
On the one hand, to improve timing precision at the longest timescale, it is
very important to have a comprehensive understanding of the sources (e.g., radiometer,
pulse phase jitter, diffractive interstellar scintillation) and the characteristics of
the noise in TOAs, and identify mitigation methods to reduce the noise 
\citep{2010arXiv1010.3785C, 2015JPhCS.610a2019W}.
On the other hand, many data analysis methods designed for detecting the weak GW signals
\citep{2010arXiv1008.1782C, 2012PhRvD..85d4034B, 2012ApJ...756..175E, 
2014ApJ...795...96W, 2015ApJ...815..125W, 2015MNRAS.449.1650Z, 
2016MNRAS.461.1317Z} 
are usually geared to work well for the data owning some specific statistical properties.
Blindly applying these data analysis strategies and pipelines without checking the
presumptions may lead to nonsensical results \citep{2016MNRAS.455.4339T}. 
In this paper, as a first step of noise characterization, we
implement a suite of robust non-parametric statistical tools to test the most important
noise properties, namely the whiteness and the Gaussianity, of the
NANOGrav 5-year (2005--2010) data published in \citet{2013ApJ...762...94D}.

Using these tools, we found that most of the frequency separated 
data are individually consistent
with the whiteness assumption, except the high frequency data from PSR J2145--0754
and J2317+1439 which show mild deviations. However, combining the data from different
frequencies as one set causes significant deviations for PSR J0613–-0200, J1455--3330, 
J1744--1134, J1909--3744, J1918--0642 and J2317+1439.  We found that this may 
be due to the minute inaccuracy of DM estimation for these pulsars with the current 
observation strategy.  For the Gaussianity, most of the data show different levels of
deviation, however, removing outliers in some pulsars would reduce the deviations.

The rest of the paper is organized as follows. In Sec.~\ref{data}, a brief description of the observation 
and the data set is given. We use the zero-crossing method to test the whiteness of the data in
Sec.~\ref{white}, and use the descriptive statistics and the hypothesis testings to check the
Gaussianity in Sec.~\ref{gauss}. Demonstrations of these analyses on 3 pulsars 
are presented in Sec.~\ref{sum}. The paper is concluded
in Sec. \ref{disc}.

\section{Observations and data}\label{data}

The NANOGrav collaboration has conducted observations with the Arecibo Observatory (AO) 
and the Green Bank Telescope (GBT), the two largest single dish radio telescopes to date. 
Currently 37 MSPs \citep{2015ApJ...813...65T} 
have been regularly timed by the NANOGrav. The first five years data (2005--2010) for 17 of
the MSPs along with the upper limit on the GW stochastic background have been published
in \citet{2013ApJ...762...94D}. In order to precisely analyze the time dependent
dispersion measure (DM) and frequency dependent pulse shape, two receivers operating
at 1.4 GHz and 430 MHz for AO and 1.4 GHz and 820 MHz for GBT have been used in
most of the observations. Observations using two different receivers were not
simultaneous. At AO, the observations from the two bands were obtained within
1 hour; whereas at GBT, the separation could be up to a week. All observations
during this 5 years have been carried out by the identical pulsar backends,
i.e. the Astronomical Signal Processor (ASP) at AO and the Green Bank Astronomical
Signal Processor (GASP) at GBT, in which the input analog signal is split
into 32 4 MHz channels (sub-bands). Due to the limitation of the real-time
computation load or the receiver bandpass, typically 16 channels would be processed
in most observations.  
The cadence between observation sessions is typically 4-6 weeks. There is a gap in the
observations of all pulsars in 2007 due to the maintenance at both telescopes.

The data product from an observation epoch is the pulse time of arrival (TOA) which
is the time of the radio emission from a fiducial rotation phase of the pulsar arriving
at the telescope. The standard TOA estimation includes polarization calibration,
pulse profile folding, profile template creation, and TOA measurement by correlating
folded profile and profile template. Those steps are integrated in the
package PSRCHIVE \citep{2004PASA...21..302H} and ASPFitsReader \citep{ferdman.phd}.
Both packages are used for cross-checking of errors which otherwise would hardly
be targeted. 

The next step of timing analysis is to fit the observed TOAs of each pulsar
to a timing model. The timing model contains a set of physical parameters
which account for the pulsar's rotation (spin period, spin period derivatives), astrometry
(position, proper motion), interstellar medium (dispersion measure), binary 
orbital dynamics, etc. This procedure is executed in the standard timing analysis
package TEMPO2 \citep{2006MNRAS.369..655H, 2006MNRAS.372.1549E} via a weighted
least square fitting. The so-called post-fit timing residuals are the differences between the measured TOAs
and the TOAs predicted by this model. A positive residual means that the observed
pulse arrives later than expected. The timing residuals potentially
contain the stochastic noise from various sources and the physical effects that
are not included in the timing model. One can refer to \citet{2013ApJ...762...94D}
for a thorough account on the NANOGrav observation strategy and timing analysis.

We can generate multiple timing residuals, denoted as $r(t,\nu)$,  from the timing 
analysis of the NANOGrav data set, where $t$ is the time of observation of a pulse in 
the Modified Julian Date (MJD), $\nu$ is the central frequency of a channel.
To simplify the study of timing effects induced by achromatic physical processes
(e.g. gravitational wave), we can average the timing residuals from
the TOAs registering the same rotation phase of the pulsar. If there are only TOAs
from one pulsar rotation in an observation epoch (true for most observations),
this averaged residual will equal to the daily averaged residual used in Fig.~1 of
\citet{2013ApJ...762...94D} and in \citet{2013arXiv1311.3693P}. 
The averaged residual $r_{I}$ for the $I$-th observation epoch in the data of
a pulsar is 
\begin{equation}\label{eq:dresidual}
r_I=\frac{\sum\nolimits_{i=1}^{N_{I}} r_{Ii}\sigma_{Ii}^{-2}}{\sum\nolimits_{i=1}^{N_{I}}\sigma_{Ii}^{-2}} \,,
\end{equation}
where $r_{Ii}=r(t_I,\nu_i)$ is the post-fit multi-frequency timing residual from the $i$-th
frequency channel at the $I$-th observation epoch, $N_{I}$ is the number of frequency
channels, and $\sigma_{Ii}$ is the uncertainty for the corresponding TOA. The
uncertainty associated with the averaged residual is
\begin{equation}\label{eq:uncert2}
\sigma_{I}=\sqrt{\left(\sum\nolimits_{i=1}^{N_{I}}\sigma_{Ii}^{-2}\right)^{-1} \frac{1}{N_{I}-1}
\sum\nolimits_{i=1}^{N_{I}}(r_{Ii}-r_{I})^{2}\sigma_{Ii}^{-2}} \,.
\end{equation}
Eq.~\ref{eq:uncert2} is the standard deviation of Eq.~\ref{eq:dresidual} with
correction of underestimation of errors in TOA. 
This estimator is suitable when $\sigma_{Ii}$  does not include all the 
noise sources associated with TOAs. 
In addition, since we have used two
independent receivers not simultaneously, we will separate the low frequency and 
high frequency averaged residuals and test them independently in our analysis.

The averaged timing residuals can be used as inputs of the GW detection pipelines.
One advantage of averaging is that it reduces the random noise components across
different frequency channels while keeps the achromatic GW signals intact.
Besides, the averaged residuals provide a quantitative way to compare with the data from
the EPTA \citep{2010CQGra..27h4014F, 2015MNRAS.453.2576L} and the PPTA 
\citep{2013PASA...30...17M}, which have routinely produced a single TOA per observation epoch.

\section{Whiteness test}\label{white}

In this section, we test the consistency of the averaged timing
residuals with the white noise assumption for each pulsar.
The white noise time series is statistically uncorrelated in time, while the distribution
of its values does not necessarily adhere to any specific probability distribution (Gaussian, Poisson, etc.).
If evenly sampled, we can use Fourier analysis to calculate the power spectrum of
the time series and to check whether it is consistent with a flat spectrum in the interested
frequency range. However the pulsar timing data are usually not evenly sampled, i.e. the
observation cadence varies, so that this conventional spectral analysis is not applicable.
The Lomb-Scargle periodgram \citep{1976Ap&SS..39..447L, 1982ApJ...263..835S}
which is designed for unevenly sampled data suffers from occasional large gaps
between observations (see Fig.~1 in \citet{2013ApJ...762...94D}), as well as limited
data volume for each pulsar (see Table~\ref{tab:Gaussianity} for detailed numbers).

It turns out that after subtracting the mean value the number of the zero-crossing 
$Z_W$ of a white noise time series is a Gaussian random variable,
\begin{equation}\label{eq:GRV}
Z_W\sim \mathcal{N}(\mu_{Z_W},\sigma_{Z_{W}}^2) \,,
\end{equation}
where the expected value $\mu_{Z_{W}}=(N-1)/2$ and the standard deviation
$\sigma_{Z_{W}}=\sqrt{N-1}/2$. The zero-crossing test is to check how large the
number of the zero-crossing for a time series is different from the expectation. It is
designed in the time domain, thus applicable to unevenly sampled data with gaps.
This test is not sensitive to any non-stationarity of the statistics of the white noise, such
as the case where the white noise has a jump in variance at some epoch because of a
change in instrumentation. It is implicit that the white noise is ``dense" which means that
all data values are non-zero and stochastic \citep{1984prvs.book.....P}.  Other kinds 
of white noise, such as the shot noise with a low shot rate will not conform to the 
zero-crossing test described here.

In Table \ref{tab:whiteness}, we show the results of the zero-crossing test
for the frequency separated averaged residuals as well as the total averaged
residuals (by combining the high and low frequency averaged residuals and sorting
them in the ascending order of corresponding TOAs) for 17 pulsars.
$\text{N}_{\text{crs}}$, is the actual number of the zero-crossing
for the data, $\Delta$ is the difference between $\mu_{Z_W}$ and $\text{N}_{\text{crs}}$.
The significance of the test is measured by how large $\Delta$ is
comparing with $\sigma_{Z_{W}}$. If $|\Delta|<2 \sigma_{Z_{W}}$ ($>$5\% in p-value 
\footnote{p-value gives the probability of obtaining a test statistic ($\text{N}_{\text{crs}}$) 
at least as extreme as the one that was actually observed, assuming that
the presumption (e.g., whiteness) is true. }),
the data is said consistent with white noise (labeled by `Y');
if $3 \sigma_{Z_{W}}>|\Delta|>2 \sigma_{Z_{W}}$, it is said mildly deviated from white 
noise (`n'); and if $|\Delta|>3 \sigma_{Z_{W}}$, it is said strongly deviated from white 
noise (`N'). We defer the detailed discussion and possible interpretation
of these results to Sec.~\ref{sum} to consolidate with the results from the Gaussianity tests.

\section{Gaussianity test} \label{gauss}

In this section, 
we first use the descriptive statistics, namely the histogram and the Q-Q plot, 
to visually inspect the general
features of the data. Then we implement a suite of inferential statistical tests
to quantitatively measure the deviations from the Gaussianity. The observation
conditions during the 5 years had been changing due to a number of factors, for instance,
radiometer noise, interstellar scintillation and instrument. Therefore the underlying
noise random variables at different frequencies and epochs are heteroscedastic.
These changes are reflected in the variations of the error bars (e.g., for the averaged 
residuals showed in Fig.~\ref{fig:0613ts}, \ref{fig:1012ts} and \ref{fig:1713ts}).
This feature is treated here by a simple normalization, so that the tested time
series are from the same underlying distribution. For the averaged
residuals, it is to normalize each residual from Eq.~\ref{eq:dresidual}
by its associated uncertainty calculated in Eq.~\ref{eq:uncert2}; While for the multi-frequency
residuals, it is to normalize each residual by the uncertainty associated with its TOA.

The inferential statistics based on the statistical hypothesis testing theory argues against
a \textit{null hypothesis} (Gaussianity) as in the mathematical proof by contradiction.
First,  the data are summarized into a single number called the \textit{test statistic},
which follows a certain probability distribution. 
Second, the \textit{p-value} are calculated based on this distribution assuming that
the null hypothesis is true. 
The lower the p-value is, the smaller the chance that the sample comes from a 
Gaussian distribution is.

One often rejects the null hypothesis when the p-value is less than the pre-assigned 
significance level  which is usually 0.05.  However, the power of the tests decreases 
as the sample size decreasing.  It is a common practice to declare the 
significance level at higher values, 
such as 0.1 or 0.2 for a small sample in order to detect possible deviation 
that may be essential.  This is an important point since the data sets 
that we will test vary largely in size (cf. Table~\ref{tab:Gaussianity}).

To avoid possible bias in different tests, the significance of the Gaussianity test is measured 
by the averaged p-value $p$ of five tests, among which the Shapiro-Wilk test (S-W) and 
the Shapiro-Francia test (S-F)  are the order statistics;  the Anderson-Darling test (A-D), 
the Cram\'{e}r-von Mises test (C-vM) and the Lilliefors test (Lillie) are base on the 
empirical distribution function (EDF). 
The results are summarized in Table~\ref{tab:Gaussianity}.   For multi-frequency residuals,
if $p>0.05$, the data is said consistent with the Gaussianity (labeled by `Y'); if $0.05>p>10^{-3}$,
the data is said mildly deviated from the Gaussianity (`n'); and if $p<10^{-3}$, the data is said strongly
deviated from the Gaussnianity (`N'). For averaged residuals, the criterion intervals are set to be
$p>0.1$ (`Y'), $0.1>p>2\times10^{-3}$ (`n'), and $p<2\times10^{-3}$ (`N'), respectively.  
The p-values of all tests are only showed for three pulsars in the legends of Fig.~\ref{fig:0613qq},
\ref{fig:1012qq}, and \ref{fig:1713qq}.

\section{Results} \label{sum}

The results for the whiteness and Gaussianity tests are summarized 
in Table~\ref{tab:whiteness} and Table~\ref{tab:Gaussianity}. 
Here, we demonstrate the tests for three pulsars in details.

\subsection{PSR J0613-0200} \label{sec:sub0613}

\begin{figure*}
\begin{center}
\includegraphics[width=1.0\textwidth]{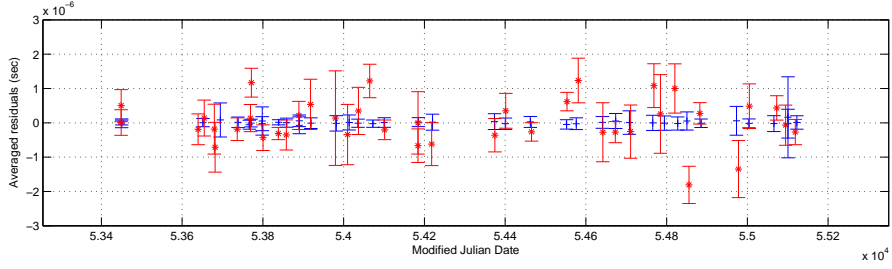}
\caption{Frequency separated averaged timing residuals with error bars for J0613--0200. The red asterisks represent high
frequency data, while the blue short-bars represent low frequency data.}
\label{fig:0613ts}
\end{center}
\end{figure*}

The frequency separated averaged timing residuals of PSR J0613-0200 are
shown in Fig.~\ref{fig:0613ts}. The red asterisks with error bars represent the
high-frequency (1.4~GHz) residuals, and blue short-bars with error bars represent the
low frequency (820~MHz) residuals. Apparently the high frequency residuals have
larger variance than the low frequency residuals, and the high frequency
error bar for this pulsar is a factor of a few larger than the low frequency
error bars. This is mainly due to the fact that the mean flux density at high
frequency is lower than that of low frequency according to the power-law 
spectrum of the flux density.  For similar integration 
time, this will result in a larger uncertainty in the measurement 
of TOA by correlating lower S/N folded pulse with 
template pulse profile \citep{1992RSPTA.341..117T}.

From Table~\ref{tab:whiteness} we can see that the low frequency
and high frequency residuals are individually consistent with the white noise
assumption. However when they are combined into a single time series, the total
residuals show more zero-crossing than expected, the deviation for this pulsar is more than
$4 \sigma$. We found 
that the excess of zero-crossing is caused by the error 
in the estimated values of time dependent DM with the observation strategy adopted in the NANOGrav 
5-year data. However, we defer the detailed discussion to Sec.~\ref{disc}.

\begin{figure*}
\begin{center}
\includegraphics[width=0.6\textwidth, angle=-90]{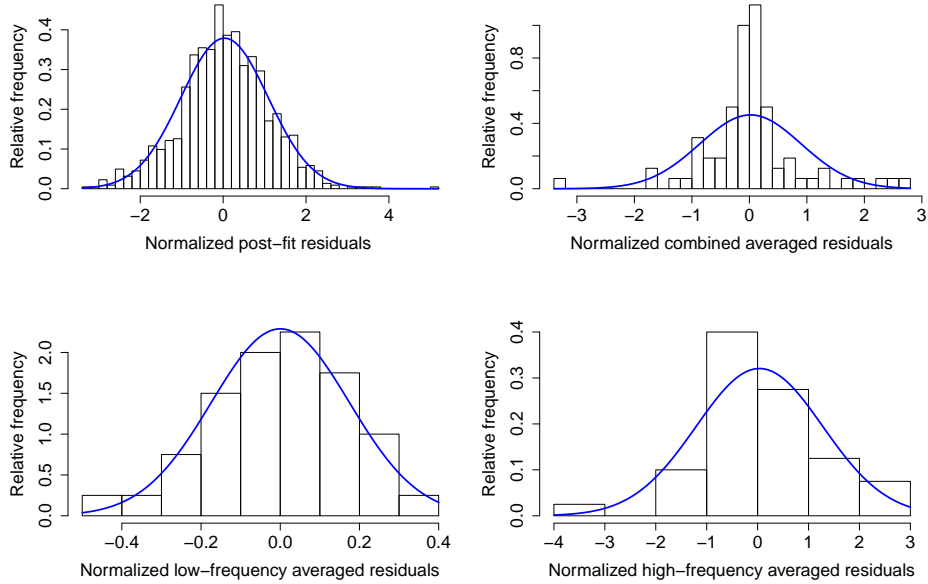}
\caption{Histograms for the post-fit multi-frequency residuals (top-left), total (top-right), 
low-frequency (bottom-left), and high-frequency (bottom-right) 
averaged residuals of J0613--0200. The blue curve is the Gaussian distribution
with the same mean and variance as the data.}
\label{fig:0613hist}
\end{center}
\end{figure*}

\begin{figure*}
\centerline{\includegraphics[width=0.6\textwidth, angle=-90]{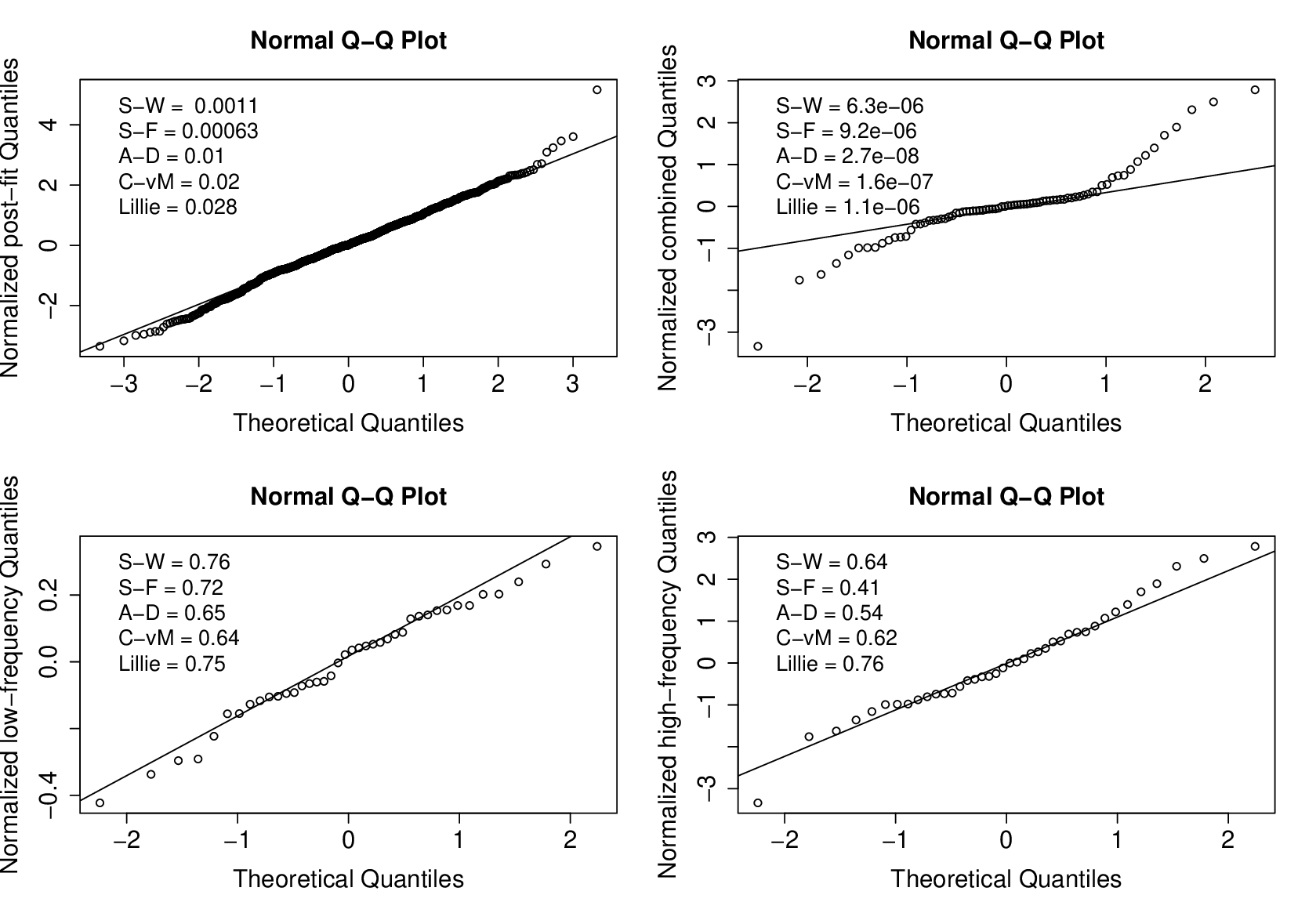}}
\caption{Quantile-Quantile plots for the post-fit multi-frequency residuals (top-left), 
total (top-right), low-frequency (bottom-left), and high-frequency  (bottom-right) 
averaged residuals of J0613--0200. The p-values of  individual test are listed in the legends.}
\label{fig:0613qq}
\end{figure*}

\begin{figure*}
\centerline{\includegraphics[width=1.0\textwidth]{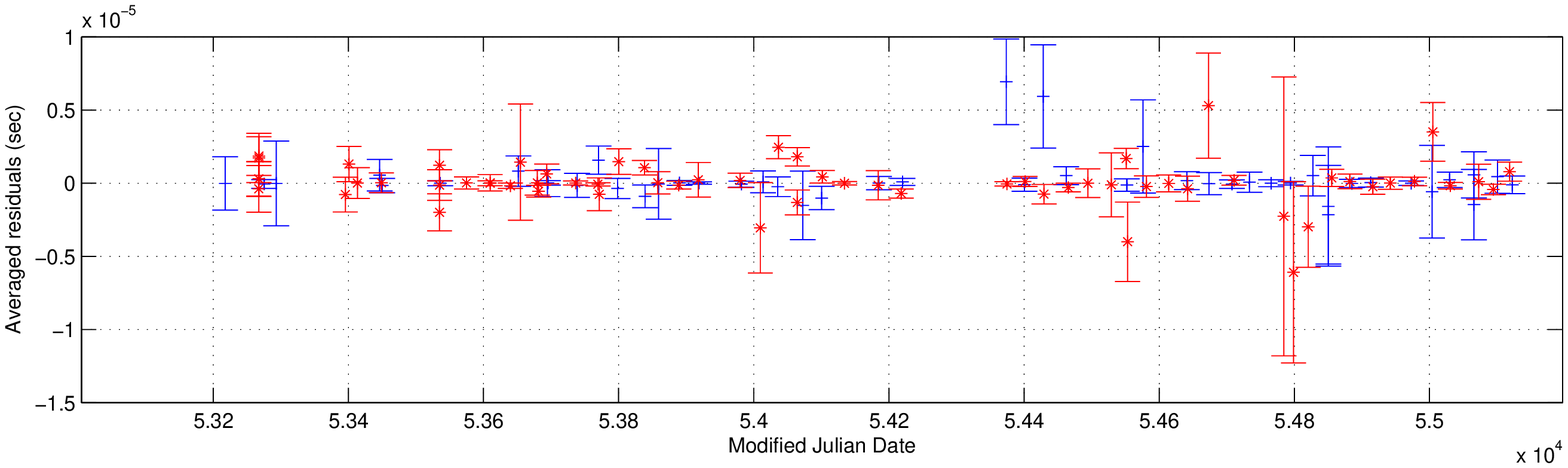}}
\caption{Frequency separated averaged timing residuals with error bars for J1012+5307. The red asterisks represent high
frequency data, while the blue short-bars represent low frequency data.}
\label{fig:1012ts}
\end{figure*}

After normalizing the averaged residuals by their associated
error bars, we notice that in Fig.~\ref{fig:0613hist} and \ref{fig:0613qq} the
standard deviation of the low frequency residuals is significantly smaller than
unity which hints that the error bar calculated for the low frequency averaged
data are overestimated. Therefore the combined residuals deviate from the
Gaussian distribution, even if the low frequency averaged residuals and high frequency
averaged residuals are both consistent with the Gaussian distribution individually. It may
suggest that in order to properly combine the data from different frequency bands in
GW detection algorithms, we may need to add a scaling parameter for each frequency band
that is similar to the EFAC\footnote{A multiplication factor for TOA error
bars of each pulsar.} parameter used in timing analysis.
The post-fit multi-frequency residuals are mildly deviated from the Gaussian distribution.

\subsection{PSR J1012+5307} \label{sec:sub1012}

From Table \ref{tab:whiteness}, we can see that the low-frequency, high-frequency,
and total averaged residuals are all consistent with the white noise assumption.
From Table.~\ref{tab:Gaussianity}, we can see that the high frequency averaged and post-fit
multi-frequency residuals are mildly deviated from the Gaussian distribution, while
the low frequency averaged and total averaged residuals are strongly
deviated from the Gaussian distribution.

\begin{figure*}
\centerline{\includegraphics[width=0.6\textwidth, angle=-90]{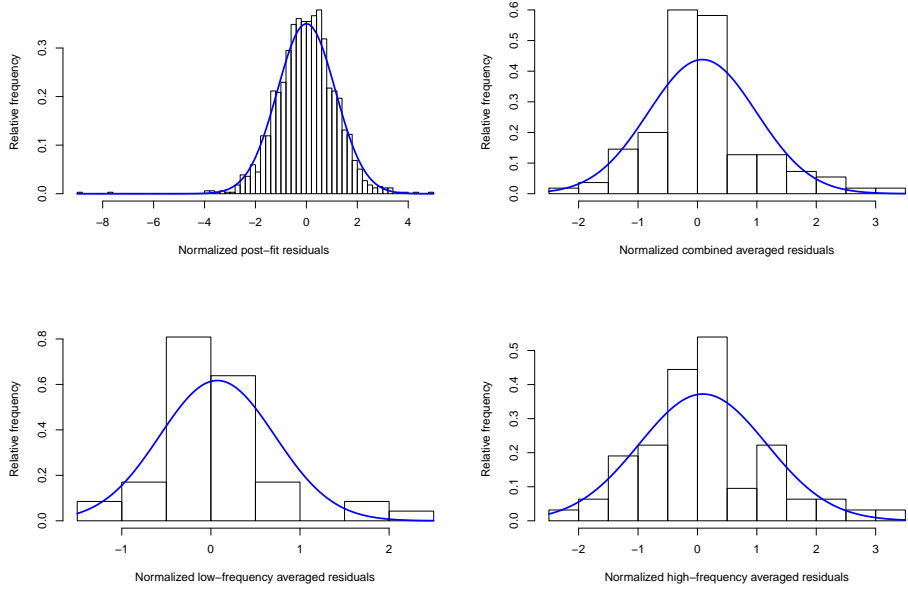}}
\caption{Histograms for the post-fit residuals, total, low-frequency, and high-frequency
averaged residuals of J1012+5307. The blue curve is the Gaussian distribution
with the same mean and variance as the data.}
\label{fig:1012hist}
\end{figure*}

\begin{figure*}
\centerline{\includegraphics[width=0.6\textwidth, angle=-90]{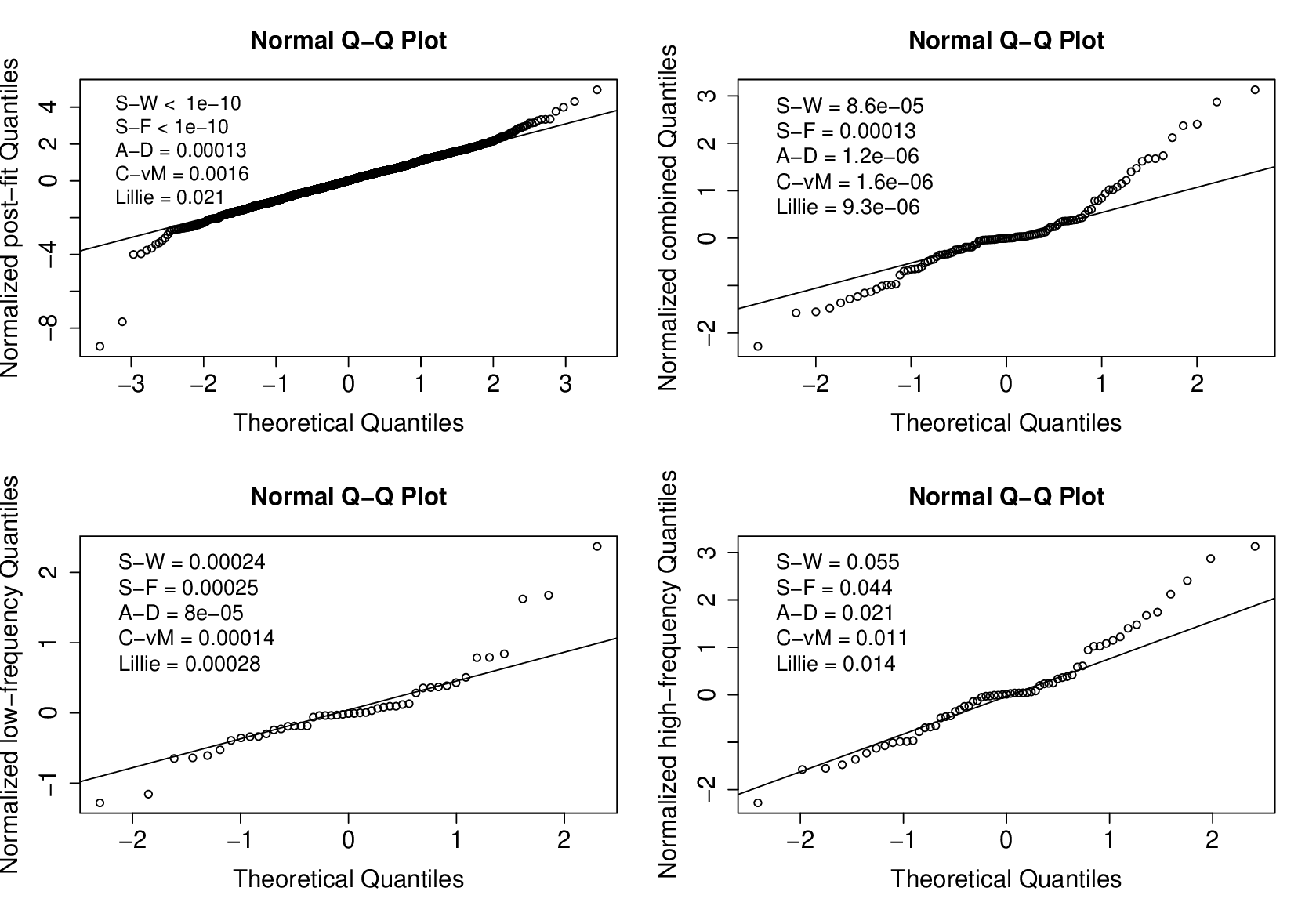}}
\caption{Quantile-Quantile plots for the post-fit residuals, total, low-frequency, and
high-frequency averaged residuals of J1012+5307. 
The p-values of  individual test are listed in the legends.}
\label{fig:1012qq}
\end{figure*}

We notice from Fig.~\ref{fig:1012qq} that for the post-fit residuals the
results from the order statistics tests (strong deviation)
are not consistent with the EDF tests (mild deviation). This is ascribed to that
the order statistic tests are sensitive to the outliers which can
be identified from Fig.~\ref{fig:1012hist} and \ref{fig:1012qq}. After
removing two outliers in the residuals, the results from the order statistics
are improved rapidly (S-W = $8.3\times10^{-4}$, S-F = $4.5\times10^{-4}$),
and become more consistent with the other tests.

\subsection{PSR J1713+0747} \label{sec:sub1713}

\begin{figure*}
\centerline{\includegraphics[width=1.0\textwidth]{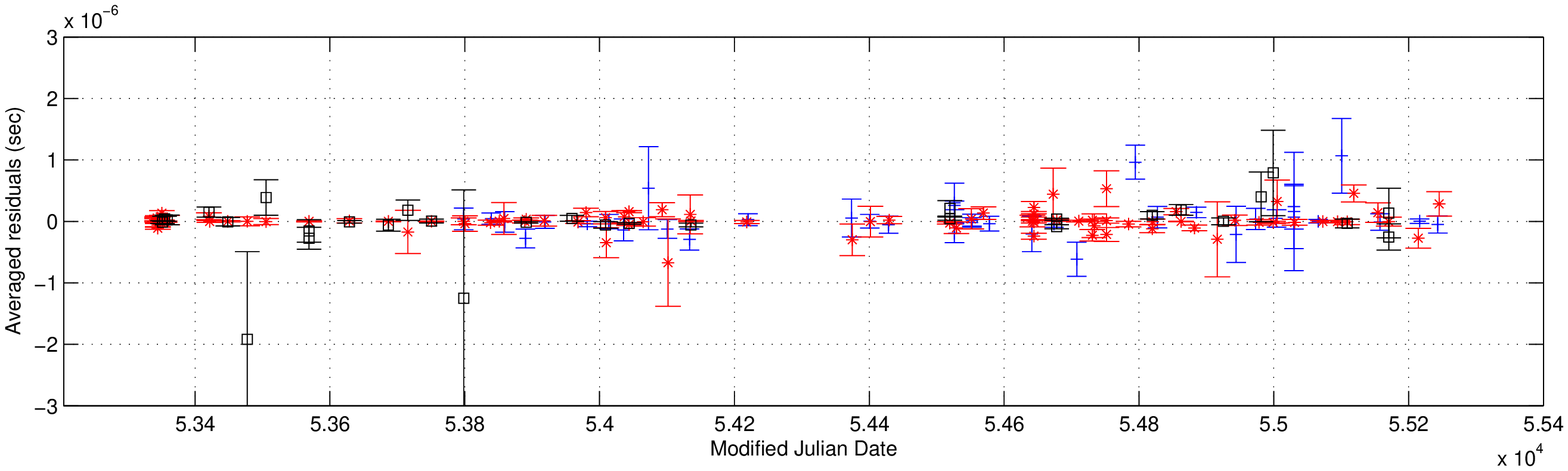}}
\caption{Frequency separated averaged timing residuals with error bars for J1713+0747. The blue short-bars represent 820 MHz
data, the red asterisks represent 1.4 GHz data, and black squares represent 2.3 GHz data.}
\label{fig:1713ts}
\end{figure*}

PSR J1713+0747 is the only one among the 17 pulsars that has been observed by both
the AO and the GBT. Currently, it is the best timed pulsar in the NANOGrav.
It has been observed extensively in three frequency bands, i.e. 820 MHz, 1.4 GHz,
and 2.3 GHz, which are marked by blue short-bars, red asterisks, and black squares respectively
in Fig.~\ref{fig:1713ts}.  (There are actually two sessions conducted in 2.7 GHz at
the AO which are not included in this analysis.)

The three frequency separated averaged residuals are all consistent with the
white noise assumption individually, whereas the total averaged residuals
are mildly deviated from it. Except the residuals from 2.3 GHz band, the residuals 
from the other two bands are all strongly deviated from the Gaussian distribution. 
And removing a few outliers improves the statistics considerably. 

\begin{figure*}
\centerline{\includegraphics[width=0.6\textwidth, angle=-90]{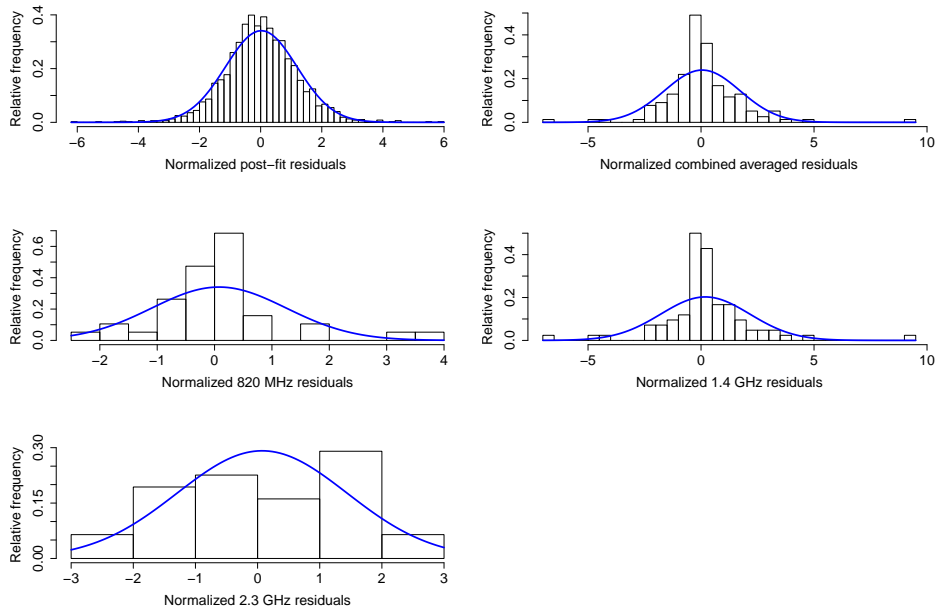}}
\caption{Histograms for the post-fit residuals, total, 820 MHz, 1.4 GHz, and 2.3 GHz
averaged residuals of J1713+0747. The blue curve is the Gaussian distribution
with the same mean and variance as the data.}
\label{fig:1713hist}
\end{figure*}

\begin{figure*}
\centerline{\includegraphics[width=0.6\textwidth, angle=-90]{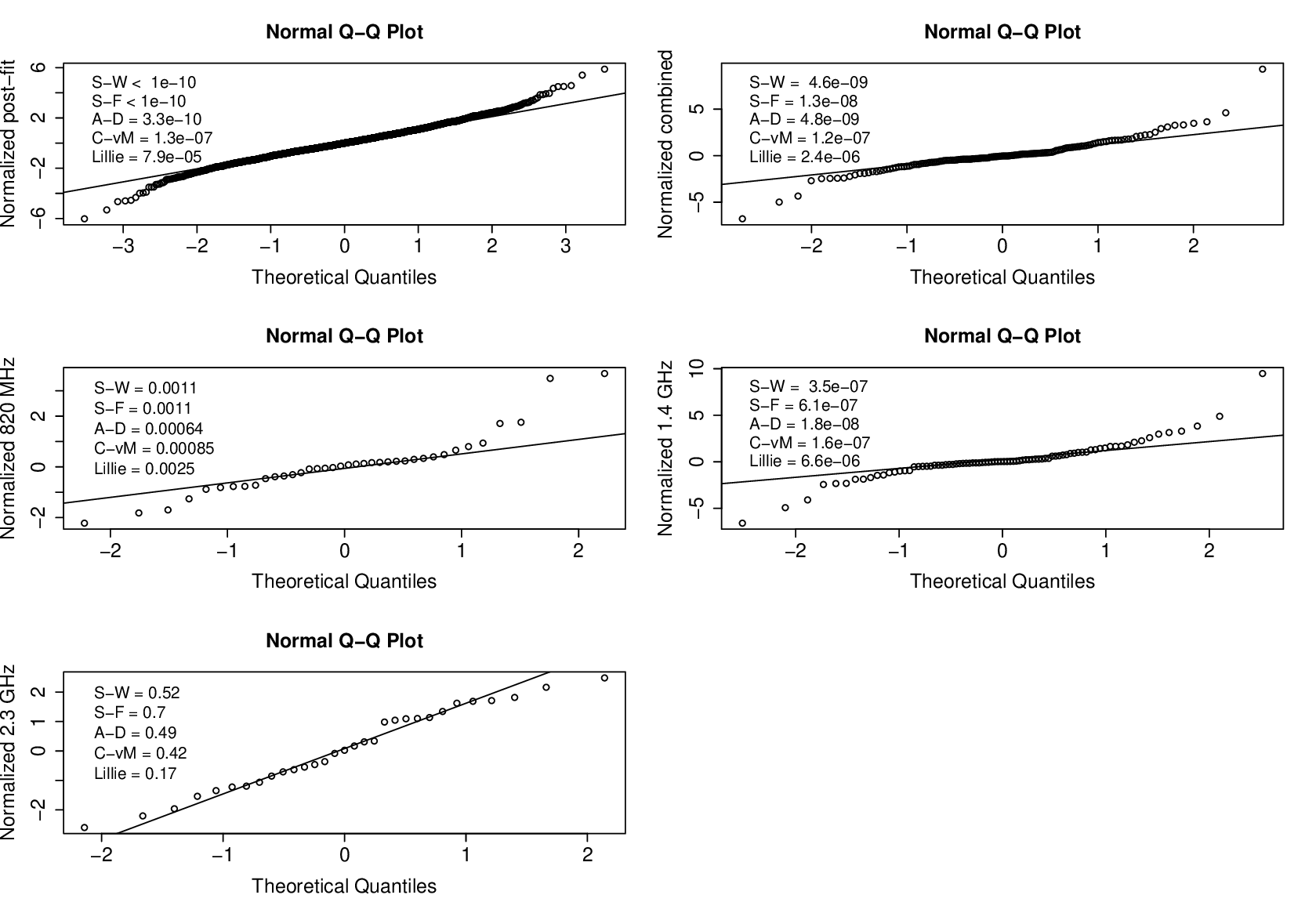}}
\caption{Quantile-Quantile plots for the post-fit residuals, total, 820 MHz, 1.4 GHz, and 2.3 GHz
averaged residuals of J1713+0747. The p-values of  individual test are listed in the legends.}
\label{fig:1713qq}
\end{figure*}

\section{Summary and discussions}\label{disc}

In this paper we utilized a set of non-parametric statistical tests to analyze the NANOGrav
5-year timing residuals for 17 pulsars. The zero-crossing has been used to test
the whiteness assumption for the averaged timing residuals. The results are summerized 
in Table~\ref{tab:whiteness}. Both descriptive 
and inferential statistical methods have been used to test
the Gaussianity for the post-fit multi-frequency and averaged timing residuals. 
The results are summarized in Table~\ref{tab:Gaussianity}. The histogram and Q-Q plots 
for 3 pulsars are shown for demonstration purpose.  

We found that for most cases,
except the high frequency averaged residuals of J2145-0750 and J2317+1439, the frequency separated
averaged residuals are consistent with the white noise assumption. However, when they are combined, 
the total averaged residuals of 5 pulsars show strong deviations from whiteness (`N') and 4 
pulsars show mild deviations from whiteness (`n'). 

In principle, the total averaged residuals can be modeled by combining two time series 
$x_1(t_i)$ and $x_2(t'_j)$ representing the low frequency and high frequency averaged 
residuals, where $t_i$ ($i=1,2,3,...,N_1$) and $t'_j$ ($j=1,2,3,...,N_2$) are not necessarily identical
or evenly spaced. If the two time series are separately drawn
from white noise processes, then the number of zero-crossing of the combined time series
(sorted in the ascending order of the union of $\lbrace t_i\rbrace$ and $\lbrace t'_j\rbrace$)
is a Gaussian random variable with the expectation equals to $(N1+N2-1)/2$ and the 
variance equals to $(N1+N2-1)/4$.

\begin{figure}
\centerline{\includegraphics[width=0.55\textwidth]{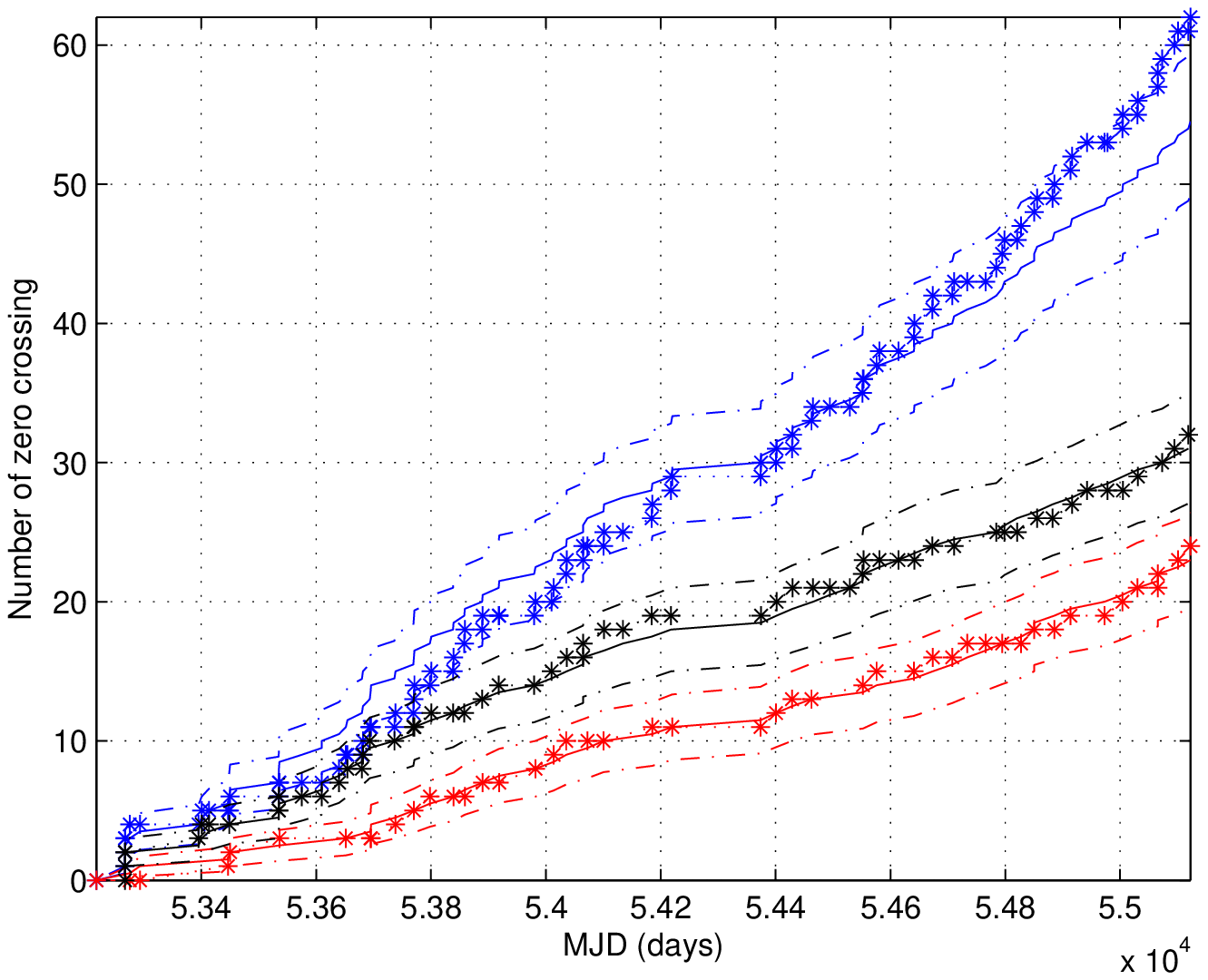}}
\caption{ Cumulative zero crossing of PSR J1012+5307, y-axis is the number of zero-crossing,
x-axis is the modified Julian date at which the observations are made. See text for details.}
\label{fig:1012zc}
\end{figure}

\begin{figure}
\centerline{\includegraphics[width=0.55\textwidth]{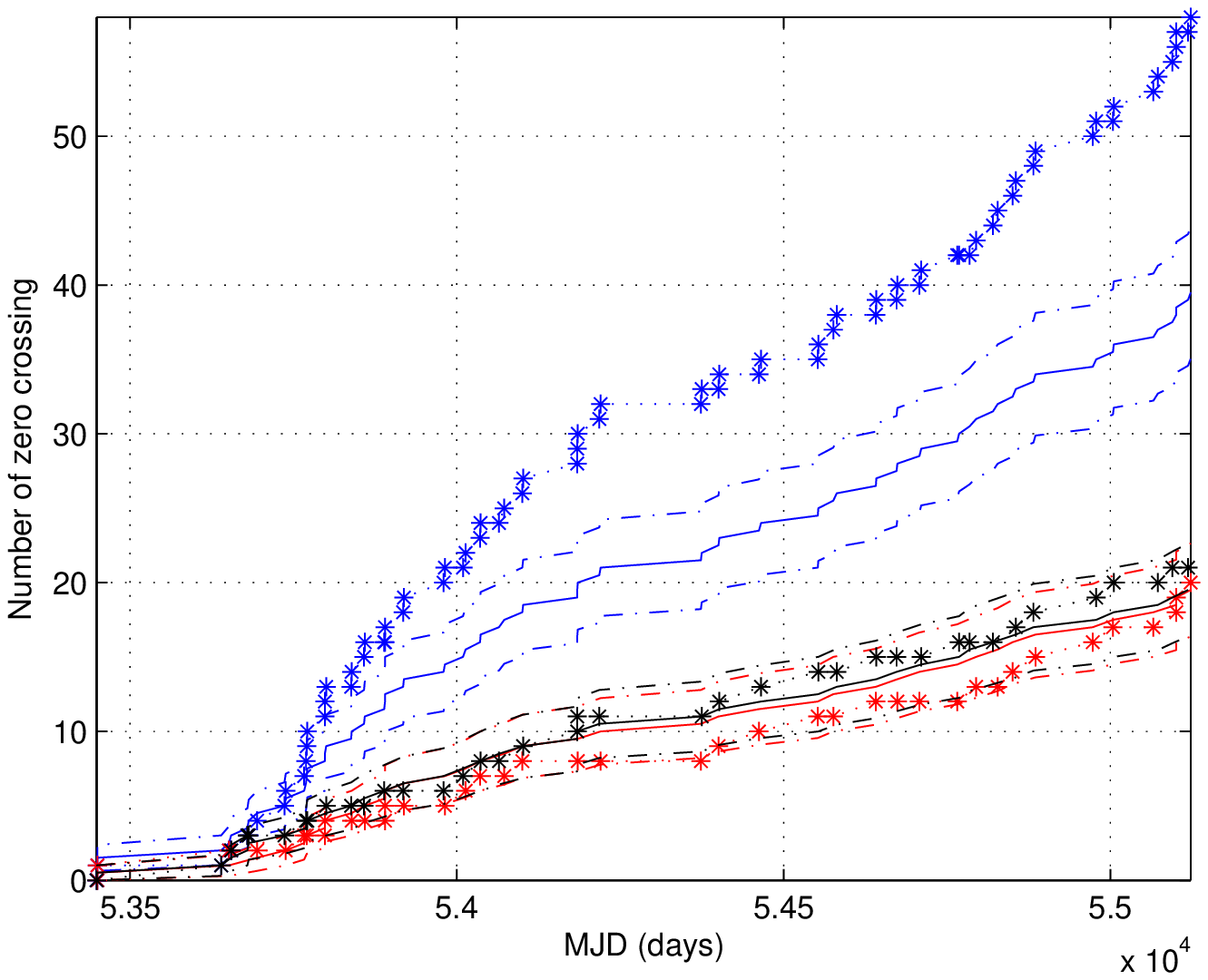}}
\caption{ Cumulative zero crossing of PSR J0613--0200, y-axis is the number of zero-crossing,
x-axis is the modified Julian date at which the observations are made. See text for details.}
\label{fig:0613zc}
\end{figure}

The cumulative zero-crossings of total averaged residuals with low and high frequency 
averaged residuals for PSR J1012+5307 and J0613--0200 are shown in Fig.~\ref{fig:1012zc} 
and Fig.~\ref{fig:0613zc}. Asterisks represent the number of zero-crossing (y-axis) added
up to this time (x-axis), it is equivalent to the zero-crossings for the data within an enlarging time
window with the left end fixed at the beginning of the time series and the right end sliding to the time
of this data point. The solid curves are the expected numbers of zero-crossing of the
data size within the window and the dash-dot lines are the $1\sigma$ contours.
They are all monotonic functions of time. Red, black, and blue represent the low frequency,
high-frequency, and total averaged residuals.

For J1012+5307, the cumulative zero-crossings of low frequency, high frequency and 
total averaged residuals follow closely to the expected values within $1\sigma$ contour. 
This is exactly what is expected for a combination of two white noise time series. 
In contrast, for J0613--0200, although the low frequency and high frequency zero-crossings
follow closely to the expectations as in J1012+5307, the combined data shows strong deviation
from its expectation. In Fig.~\ref{fig:0613zc}, it starts to deviate away from the 
beginning of the time
series, and the final deviation is more than $4\sigma$. As stated, this strong deviation for the
combined averaged residuals also appears in other pulsars in Table~\ref{tab:whiteness}.

In fact, the apparent excess of zero-crossing is mainly due to the strategy of fitting 
the physical parameters especially the time-variable DM in timing analysis. It is the practice in the 
NANOGrav 5-year data timing analysis to include a piecewise-constant DM(t) function in the fitting model along 
with the other parameters (rotation, astrometry, binary dynamics, and pulse 
profile evolution). The window for a constant DM value is typically 15 days which 
include a couple of observations conducted at high and low frequencies. 
However, any fluctuation of DM within this window or inaccuracy of the DM 
fit will introduce additional error between the adjacent averaged timing residuals from 
two widely separated bands as follows, 
\begin{equation}\label{eq:dm2t}
\delta t\simeq 4.15 \times 10^{6} ~\text{ms} \times\delta \text{DM} \times  \left(f_{1}^{-2}-f_{2}^{-2}\right)\,.
\end{equation}
Here, $f$ is measured in MHz.   For J0613--0200, 
the uncertainty of DM measurement ($\sim 10^{-4} \rm{cm^{-3} pc}$) 
can produce several hundreds of nanosecond of fluctuation between low (820 MHz) 
and high frequency (1.4 GHz). This is comparable 
with the RMS of averaged timing residual reported in Table 2 of \citet{2013ApJ...762...94D}. 
The minute error of DM will cause low frequency TOAs tend to be advanced and high 
frequency TOAs delayed or vise versa (the DM fit will tend to move the two sets of TOAs from 
low and high frequencies, so that their averaged residual is zero) which will produce extra zero-crossings between 
low and high frequency timing residuals within a DM fitting window. 
This effect is expected to be seen more clearly in the GBT observed pulsars, since the 
time separation between two observation bands is much larger and the frequency coverage 
(crucial for the DM measurement) is significantly smaller than the AO.  
We found that all 5 pulsars that have total averaged residuals strongly deviated 
from whiteness, whereas frequency separated averaged residuals are all consistent, 
are observed by the GBT.  

The Gaussianity is one of the fundamental assumptions used in most if not all of the
GW detection methods. The tests here suggest that many of the NANOGrav pulsars show deviations
from the Gaussian distribution at different levels. 
The deviations in some data, averaged and multi-frequency post-fit residuals,
can be mitigated by removing a few outliers. This strategy is consistent with the so called robust statistics
\citep{2002PhRvD..65l2002A, 2003PhRvD..67l2002A} which is used to confront the
non-Gaussianity in GW data analysis by clipping samples with values locating in the 
outlier part of the probability distribution.  It is robust in the sense that it is close to optimal for the Gaussian
noise but far less sensitive to the large excess events than the conventional statistics.
Moreover, for the purpose of detection, coherent methods 
(e.g., \citet{2014ApJ...795...96W, 2015ApJ...815..125W}) have been
shown to be robust against non-Gaussianity for detecting deterministic GW signals \citep{2001PhRvD..63j2001F}.
Alternative method in wavelet domain has also been explored for searching stochastic
GW signals in non-Gaussian and non-stationary noise \citep{2002gr.qc.....8007K} 
with ground based GW detectors.  The results here suggest that these methods 
should be investigated for the GW detection with PTA in the future.

\addtocounter{footnote}{1} 
\footnotetext[\value{footnote}]{High frequency observations from 2.3 GHz, see Sec.~\ref{sec:sub1713} for details. }   

\begin{table*}
\caption{Results for the zero-crossing test. Consistent with whiteness -- `Y',
mildly deviated -- `n', 
and strongly deviated -- `N'. }  
\label{tab:whiteness}
\begin{center}
\begin{tabular}{c|ccccc|ccccc|ccccc}
\hline\hline
Source & \multicolumn{5}{|c|}{Low-frequency band} &
\multicolumn{5}{|c|}{High-frequency band} & \multicolumn{5}{|c}{Combined} \\
& $\text{N}_{\text{crs}}$ & $\mu_{Z_W}$ & $\Delta$ & $\sigma_{Z_{W}}$ & Y/n/N
& $\text{N}_{\text{crs}}$ & $\mu_{Z_W}$ & $\Delta$ & $\sigma_{Z_{W}}$ & Y/n/N
& $\text{N}_{\text{crs}}$ & $\mu_{Z_W}$ & $\Delta$ & $\sigma_{Z_{W}}$ & Y/n/N \\
\hline
J0030+0451 & 12 & 11.5  & -0.5  & 2.4 & Y
           & 14 & 12.5  & -1.5  & 2.5 & Y
           & 27 & 24.5  & -2.5  & 3.5 & Y \\

J0613--0200 & 20 & 19.5 & -0.5  & 3.1 & Y
            & 19 & 19.5 &  0.5  & 3.1 & Y
            & 58 & 39.5 & -18.5 & 4.4 & N \\

J1012+5307 & 26 & 23   & -3   & 3.4 & Y
           & 34 & 31   & -3   & 3.9 & Y
           & 62 & 54.5  & -7.5  & 5.2  & Y \\

J1455--3330 & 22 & 20  & -2 & 3.2 & Y
            & 23 & 22  & -1 & 3.3 & Y
            & 58 & 42.5 & -15.5 & 4.6 & N \\

J1600--3053 & 13 & 10.5 & -2.5 & 2.3 & Y
            & 13 & 12.5 & -0.5 & 2.5 & Y
            & 33 & 23.5 & -9.5 & 3.4 & n \\

J1640+2224 & 18 & 16.5 & -1.5 & 2.9 & Y
           & 14 & 15.5 & 1.5  & 2.8 & Y
           & 35 & 32.5 & -2.5 & 4.0 & Y \\

J1643--1224 & 26 & 23   & -3   & 3.4 & Y
            & 22 & 24.5 & 2.5  & 3.5 & Y
            & 62 & 48   & -14  & 4.9 & n \\

J1713+0747  & 19 & 18.5 & -0.5 & 3 & Y
           & 46 & 41.5 & -4.5 & 4.6 & Y
           & - & - & - & - & - \\

J1713+0747 
           & - & - & - & - & -
           & 18 & 15 & -3 & 2.7 & Y
           & 94 & 76 & -18 & 6.2 & n \\

J1744--1134 & 25 & 23.5  & -1.5  & 3.4 & Y
            & 33 & 29.5  & -3.5  & 3.8 & Y
            & 70 & 53.5  & -16.5 & 5.2 & N \\

J1853+1308 & 21 & 20 & -1 & 3.2 & Y
           & -  & -  & -  & -   & -
           & 19 & 21 & 2  & 3.2 & Y \\

B1855+09 & 17 & 18   & 1    & 3.0  & Y
         & 17 & 15.5 & -1.5 & 2.8  & Y
         & 46 & 34   & -12  & 4.1  & n \\

J1909--3744 & 17 & 17   & 0     & 2.9 & Y
            & 17 & 16   & -1    & 2.8 & Y
            & 52 & 33.5 & -18.5 & 4.1 & N \\

J1910+1256 & 13 & 15  & 2    & 2.7 & Y
           & 3  & 2.5 & -0.5 & 1.1 & Y
           & 21 & 18  & -3   & 3   & Y \\

J1918--0642 & 20 & 19.5 & -0.5  & 3.1  & Y
            & 29 & 26.5 & -2.5  & 3.6  & Y
            & 64 & 46.5 & -17.5 & 4.8  & N \\

B1953+29 & 14 & 11  & -3   & 2.3 & Y
         & -  & - & - & - & -
         & 16 & 12  & -4   & 2.4 & Y \\

J2145--0750 & 15  & 10.5  & -4.5  & 2.3 & Y
            & 18  & 11.5  & -6.5  & 2.4 & n
            & 31  & 22.5  & -8.5  & 3.4 & n \\

J2317+1439 & 26 & 21   & -5    & 3.2 & Y
           & 27 & 20   & -7    & 3.2 & n
           & 56 & 41.5 & -14.5 & 4.6 & N \\
\hline\hline
\end{tabular}
\end{center}
\end{table*}

\begin{table*}
\caption{Results for the inferential statistical tests of the Gaussianity.
The numbers represent the sample size.  For post-fit residuals,
if $p>0.05$, the data is consistent with Gaussianity (labeled by `Y'), if $0.05>p>10^{-3}$,
the data is mildly deviated from Gaussianity (`n'), and if $p<10^{-3}$, the data is strongly
deviated from Gaussnianity (`N'). For averaged residual, the criterion intervals are set to be
$p>0.1$ (`Y'), $0.1>p>2\times10^{-3}$ (`n'), and $p<2\times10^{-3}$ (`N') respectively.
`NA' appears when the test is not applicable to such small sample size.}
\label{tab:Gaussianity}
\begin{center}
\begin{tabular}{c|rr|r|r|r|r|r|r|c}
\hline\hline
Source & $P$ & DM
  & \multicolumn{6}{|c|}{Averaged timing residuals}
  & Post-fit  \\
  & (ms) & (pc~cm$^{-3}$)
  &  327 MHz & 430 MHz & 820 MHz & 1.4 GHz & 2.3 GHz & Comb.
  &  \\
\hline
J0030+0451 & 4.87 & 4.33  & - & 24 N & - & 26 n & - & 50 N & 545 Y \\
J0613--0200 & 3.06 & 38.78  & - & - & 40 Y & 40 Y & - & 80 N & 1113 n \\
J1012+5307 & 5.26 & 9.02 & - & - & 47 N & 63 n & - & 110 N & 1678 n \\
J1455--3330 & 7.99 & 13.57 & - & - & 41 n & 45 Y & - & 86 N & 1100 n \\
J1600--3053 & 3.60 & 52.33 & - & - & 22 n & 26 n & - & 48 n & 625 Y \\
J1640+2224 & 3.16 & 18.43 & - & 34 N & - & 32 N & - & 68 N & 631 N \\
J1643--1224 & 4.62 & 62.42 & - & - & 47 N & 50 Y & - & 97 N & 1266 N \\
J1713+0747 & 4.57 & 15.99 & - & - & 38 N & 84 N & 31 Y & 153 N & 2368 N \\
J1744--1134 & 4.07 & 3.14 & - & - & 48 N & 60 Y & - & 108 N & 1617 N \\
J1853+1308 & 4.09 & 30.57 & - & - & - & 41 Y & 2 NA & 43 Y &  497 Y \\
B1855+09 & 5.36 & 13.30 & - & 37 N & - & 32 Y & - & 69 n & 702 N \\
J1909--3744 & 2.95 & 10.39 & - & - & 35 n & 33 Y & - & 68 N & 1001 N \\
J1910+1256 & 4.98 & 34.48 & - & - & - & 31 Y & 6 Y & 37 Y & 525 Y \\
J1918--0642 & 7.65 & 26.60 & - & - & 40 Y & 54 n & - & 94 N & 1306 Y \\
B1953+29 & 6.13 & 104.50 & - & - & - & 23 Y & 2 NA & 25 Y & 208 Y \\
J2145--0750 & 16.05 & 9.03 & - & - & 22 N & 24 Y & - & 46 n & 675 n \\
J2317+1439 & 3.45 & 21.90 & 43 n & 41 n & - & - & - & 84 N & 458 n \\
\hline\hline
\end{tabular}
\end{center}
\end{table*}

\normalem
\begin{acknowledgements}

This work was supported by the National Science Foundation under PIRE grant 0968296.
We are grateful to the NANOGrav members for helpful comments and discussions.
Y.W. acknowledges the support by the National Science Foundation of China
(NSFC) under grant numbers 11503007 and 91636111. 
D.R.M. acknowledges partial support through the New York Space Grant Consortium. 
J.A.E. acknowledges support by NASA through Einstein Fellowship grant PF4-150120.
M.V. acknowledges support from the JPL RTD program. 
Data for this project were collected using the facilities
of the National Radio Astronomy Observatory and the
Arecibo Observatory. The National Radio Astronomy Observatory
is a facility of the NSF operated under cooperative
agreement by Associated Universities, Inc. The Arecibo Observatory
is operated by SRI International under a cooperative
agreement with the NSF (AST-1100968), and in alliance with
Ana G. M\'{e}ndez-Universidad Metropolitana and the Universities
Space Research Association.

\end{acknowledgements}
  

\end{document}